\documentclass[journal,10pt]{IEEEtran}
\usepackage[T1]{fontenc}
\usepackage{cite}
\usepackage{amsthm}
\usepackage{amssymb}
\usepackage{amsfonts}
\usepackage{fancyhdr}  %
\usepackage{extarrows}
\usepackage{algorithm}
\usepackage{multirow,tabularx}
\usepackage{mathtools}
\usepackage{xcolor}
\usepackage[english]{babel}
\usepackage{caption}
\usepackage{bm}
\usepackage{algorithmic}
\usepackage{amsmath}
\usepackage{subfigure}
\usepackage{makecell}
\usepackage{colortbl}
\usepackage{booktabs}
\usepackage{amsmath}
\usepackage{graphicx}
\usepackage{diagbox}
\usepackage{cancel}
\usepackage{tablefootnote}
\usepackage{makecell}

\usepackage{hyperref} 
\usepackage{cleveref} 
\usepackage{setspace}
\hypersetup{colorlinks = true, 
	    linkcolor = black, 
	    urlcolor = black,
            citecolor = black} 

\begin{document}
\title{
Channel Mapping Based on Interleaved Learning with Complex-Domain MLP-Mixer
}
\author{Zirui~Chen,
		Zhaoyang~Zhang, 
		Zhaohui~Yang, and Lei~Liu
\thanks{This work was supported in part by National Natural Science Foundation of China under Grants 62394292 and U20A20158, Ministry of Industry and Information Technology under Grant TC220H07E, Zhejiang Provincial Key R\&D Program under Grant 2023C01021, and the Fundamental Research Funds for the Central Universities. (\textit{Corresponding Author: Zhaoyang~Zhang})}
\thanks{Z.~Chen, Z.~Zhang, Z.~Yang and L.~Liu are with College of Information Science and Electronic Engineering, Zhejiang University, Hangzhou 310027, China, and also with Zhejiang Provincial Key Laboratory of Info. Proc., Commun. \& Netw. (IPCAN), Hangzhou 310007, China. (E-mails: \{ziruichen, ning\_ming, yang\_zhaohui, lei\_liu\}@zju.edu.cn)}
}

\maketitle
\begin{abstract}
In multiple-input multiple-output (MIMO) orthogonal frequency division multiplexing (OFDM) systems, representing the whole channel only based on partial subchannels will significantly reduce the channel acquisition overhead. For such a channel mapping task, inspired by the intrinsic coupling across the space and frequency domains, this letter proposes to use interleaved learning with partial antenna and subcarrier characteristics to represent the whole MIMO-OFDM channel. Specifically, we design a complex-domain multilayer perceptron (MLP)-Mixer (CMixer), which utilizes two kinds of complex-domain MLP modules to learn the space and frequency characteristics respectively and then interleaves them to couple the learned properties. The complex-domain computation facilitates the learning on the complex-valued channel data, while the interleaving tightens the coupling of space and frequency domains. These two designs jointly reduce the learning burden, making the physics-inspired CMixer more effective on channel representation learning than existing data-driven approaches. Simulation shows that the proposed scheme brings 4.6\textasciitilde10dB gains in mapping accuracy compared to existing schemes under different settings. Besides, ablation studies show the necessity of complex-domain computation as well as the extent to which the interleaved learning matches the channel properties.
\end{abstract}
\begin{IEEEkeywords}
  Channel Mapping, Deep learning, Physics-inspired learning, MIMO, OFDM.
\end{IEEEkeywords}

\IEEEpeerreviewmaketitle

\section{Introduction}\label{section1}
Multiple-input multiple-output (MIMO) and orthogonal frequency division multiplexing (OFDM) modulation are two key enabling technologies in wireless communication systems, thanks to their ability in achieving high spectrum efficiency. 
Meanwhile, unleashing these potentials generally requires acquiring real-time channel state information (CSI). However, MIMO and OFDM techniques significantly increase the size of CSI data, resulting in significant signaling overhead  \cite{bigAI6G}.

To address this challenge, the work in \cite{channel_mapping} proposes the channel mapping over space and frequency, i.e., by leveraging the implicit correlations within the high-dimensional channel, obtaining the whole MIMO-OFDM channel from known subchannel at a partial set of antennas and subcarriers. In this way, the signaling overhead required for channel acquisition can be reduced from the entire high-dimensional channel to some specific subchannels. Specifically, the authors in \cite{channel_mapping} leverage the capabilities of deep learning (DL) techniques in mining implicit features and high-dimensional data representation, using a multi-layer perceptron (MLP) to learn the channel mapping function by fitting it from the training data. Moreover, there are some channel estimation works that use similar ideas to estimate the whole channel by mapping from only a small number of subchannels. In \cite{channelesnet}, the authors use DL learning techniques to map the information of partial subcarriers to whole subcarriers for channel estimation in OFDM systems. Further, the work in \cite{res_cnn_es} improves the network structure in \cite{channelesnet} with a residual convolutional neural network (Res\_CNN) to enhance the performance of channel mapping and estimation.

Despite some preliminary attempts, existing channel mapping networks are still directly migrated from popular DL schemes and lack task-related design, which makes the learning mainly data-driven. However, data-driven learning is mediocre under limited training data, and the learned generalization is often weak. Therefore, it is necessary to design the network a priori according to the properties of the data and the task to improve the learning efficiency, such as the design of CNNs based on the translational equivalence and smoothness of images.
This letter delves into the physical properties inside MIMO-OFDM channels and accordingly proposes a physics-inspired learning scheme for channel mapping. The main contributions of this letter are summarized as follows:
\begin{itemize}
    \item By revealing the intrinsic coupling between the space and frequency characteristics of an MIMO-OFDM channel, we propose an interleaved learning framework to investigate this cross-domain internal correlation.
    \item To facilitate the learning process, we propose a novel complex-domain MLP-Mixer (CMixer) layer module for channel representation, and then construct a CMixer model for channel mapping by stacking multiple CMixer layers.
    \item Through comparison experiments with existing methods, we demonstrate the superiority of the proposed scheme. Further ablation studies show the necessity and value of physics-inspired design for improving mapping performance.
\end{itemize}

\section{System Model}\label{system model}
In this section, we introduce the system model, including the channel model and the channel mapping framework.

\subsection{Channel Model}\label{section2.1}
We consider a massive MIMO system, where a base station (BS) with $N_\mathrm{t} \gg 1$ antennas serves multiple single-antenna users. Also, the considered system adopts OFDM modulation with $N_\mathrm{c}$ subcarriers. The channel between one user and the BS is assumed to be composed of $P$ paths, which can be expressed as,
\begin{equation}
   {\bf{h}}\left( f \right) = \sum\limits_{p = 1}^P {{\alpha _p}{e^{ - j2\pi f{\tau _p}}}{\bf{a}}(\vec p )},
\label{channel}
\end{equation}
where $f$ is the carrier frequency, ${\alpha _p}$ is the amplitude attenuation, ${\tau _p}$ is the propagation delay, and $\vec p$ is the three-dimensional unit vector of departure direction of the $p$-th path. Furthermore, $\mathbf{a}(\vec p )$ is the array defined as,
\begin{equation}
   {\mathbf{a}(\vec p )} = {\left[ {1,{e^{ - j2\pi f {{\vec d_1}}  \cdot \vec p /c}}, \ldots ,{e^{ - j2\pi f {{\vec d_{{N_{\rm{t}}} - 1}}}  \cdot \vec p /c}}} \right]^T},
\label{array_vector}
\end{equation}
where $\left[ {\vec 0 , {{\vec d_1}} , \ldots , {{\vec d_{{N_{\rm{t}}} - 1}}} } \right]$ is the space vector array, ${{ \vec d_i}}~(i=1,2,\ldots, {{N_{\rm{t}}} - 1})$ represents the  three-dimensional space vector between the $i$-antenna and the first antenna, and $c$ is the speed of light. The transmission distance shift between the $i$-th antenna and the first antenna on $p$-th path can be written as $ {{\vec d_i}}  \cdot \vec p $. If the antenna array form is a uniform linear array, ${{ \vec d_i}}  \cdot \vec p $ can be simplified as $id\cos {\theta _p}$, where ${\theta _p}$ is the angle of departure (AoD) of the $p$-th path and $d$ is the antenna space. This simplified case is equivalent to the channel model in the literature \cite{2Dseq2seq}.
Further, the whole CSI matrix $\mathbf{H} \in {\mathbb{C}^{{N_\mathrm{t}}{\rm{ \times }}{N_{\mathrm{c}}}}}$ between the user and the BS can be expressed as, 
\begin{align}
   \mathbf{H} = [\mathbf{h}({f_1}),\mathbf{h}(f_2), \cdots ,\mathbf{h}({f_{N_{\mathrm{c}}}})],
\label{eq_channel_matrix}
\end{align}
where ${f_i} = {f_0} + \Delta {f_i}, (i = 1,2, \cdots ,{N_{\rm{c}}})$, ${f_i}$ is the $i$-th subcarrier frequency, $f_0$ is the base frequency, and $\Delta {f_i}$ is the frequency shift between the $i$-th subcarrier and the base frequency. 

\vspace{-1em}
\subsection{Channel Mapping}\label{section2.2}
As shown in the channel model, there are significant correlations within the CSI matrix, benefiting from the path-similarity of sub-channels. Further, channel mapping \cite{channel_mapping} aims at realizing such a task leveraging on this correlation: inputting sub-channels of some antennas and subcarriers to output the whole channel data, which can be written as,
\begin{equation}
    {\rm{g}}:{\bf{H}}[\Omega ] \to {\bf{H}},
\label{eq_mapping}
\end{equation}
where $\Omega$ is the selected small subset in space and frequency, ${\bf{H}}[\Omega ]$ is also written as ${\bf{H}}^0$.

The channel mapping task has many potential applications. For example, in high-dimensional channel estimation, it is possible only to estimate the CSI of several antennas and subcarriers and then obtain the whole channel by channel mapping to decrease the pilot overhead. Furthermore, in the high-dimensional CSI feedback, it is possible only to upload partial downlink CSI, and the BS can still reconstruct the whole downlink CSI based on channel mapping, forming a user-friendly CSI feedback scheme without the additional encoder. In Eq. \eqref{eq_mapping}, for the subset $\Omega$, a typical selection is the whole channel of several certain antennas and subcarriers, i.e.,
\begin{align}
    \begin{array}{l}
\Omega {\rm{ = A}} \otimes {\rm{B}},
\end{array}
\end{align}
where $\rm{A}$ is the subset of the selected antennas, $\rm{B}$ is the subset of the selected subcarriers, and $\otimes$ represents that $\forall a \in {\rm{A\& b}} \in {\rm{B,}}\left( {a,b} \right) \in \Omega $. Besides, the settings of \rm{A} and \rm{B} are often uniform distributions and as widely ranging as possible in space and frequency domains, respectively.

\begin{figure}[htbp]
  \centering
  \includegraphics[width=0.96\linewidth]{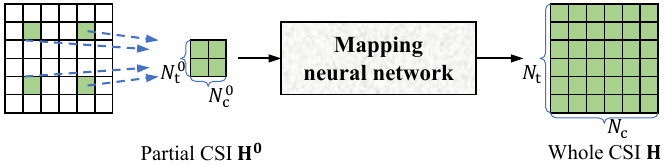}
  \caption{The overall framework of channel mapping.}
  \label{fig_channel_mapping}
  \vspace{-0.2em}
\end{figure}

Since CSI is a complex coupling of multipath channel responses, the internal correlation in the CSI matrix is often implicit and, thus, is difficult to explore through traditional signal processing methods. Therefore, we use the DL technology with excellent implicit feature learning capabilities to realize efficient channel mapping. The overall DL-based channel mapping framework is shown in Fig. \ref{fig_channel_mapping}. This learning process mainly relies on the representation learning of channel data, i.e., representing implicit features from known channel information and then representing the whole channel based on the obtained features. For the mapping neural network, if some a priori design can be introduced according to the physical properties of the MIMO-OFDM channel, it will undoubtedly improve the representation learning and guide the network to learn more efficiently from the limited training data.

\section{Proposed Schemes and Related Analysis}\label{section3}
This section details our proposed learning scheme and related analysis, including the revealed channel characteristic, the learning module inspired by this characteristic, and the proposed channel mapping scheme utilizing this module.

\begin{figure*}[t]
\centering
  \begin{minipage}[t]{0.006\textwidth}
    ~
  \end{minipage}
  \centering
  \begin{minipage}[t]{0.18\textwidth}
    \centering
    \includegraphics[width=\textwidth]{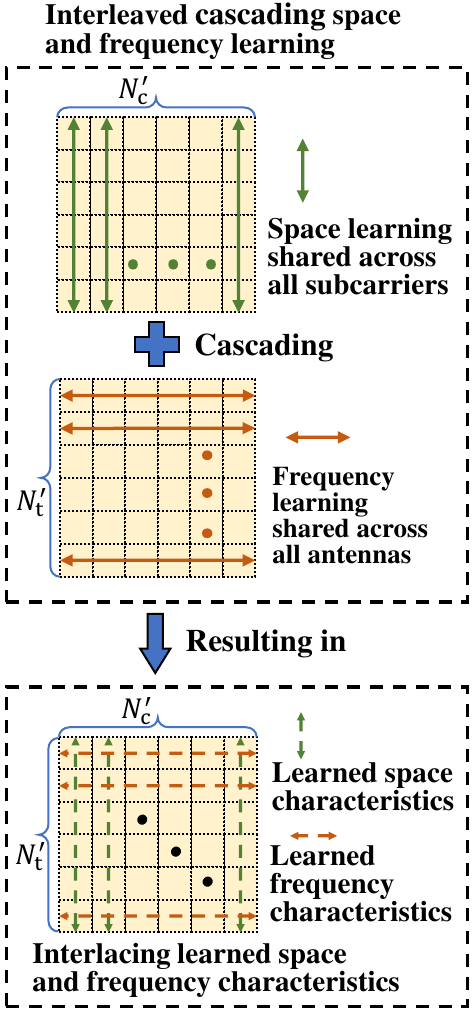}
    \caption*{\footnotesize(a) Overall idea of interleaved learning}
  \end{minipage}
  \begin{minipage}[t]{0.022\textwidth}
    ~
  \end{minipage}
  \begin{minipage}[t]{0.216\textwidth}
    \centering
    \includegraphics[width=\textwidth]{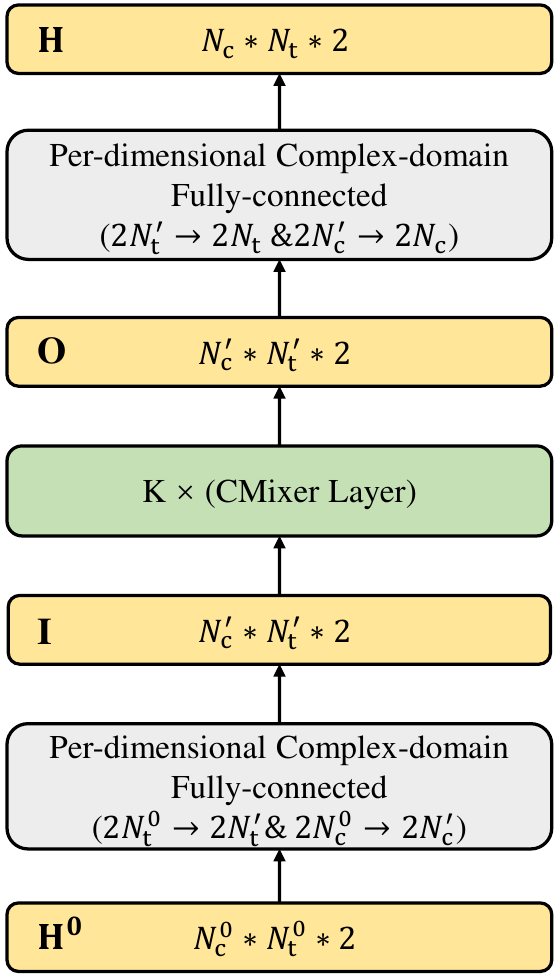}
    \caption*{\footnotesize(b) Stacking interleaved learning modules to form CMixer model}
  \end{minipage}
  \centering
  \begin{minipage}[t]{0.023\textwidth}
    ~
  \end{minipage}
  \begin{minipage}[t]{0.19\textwidth}
    \centering
    \includegraphics[width=\textwidth]{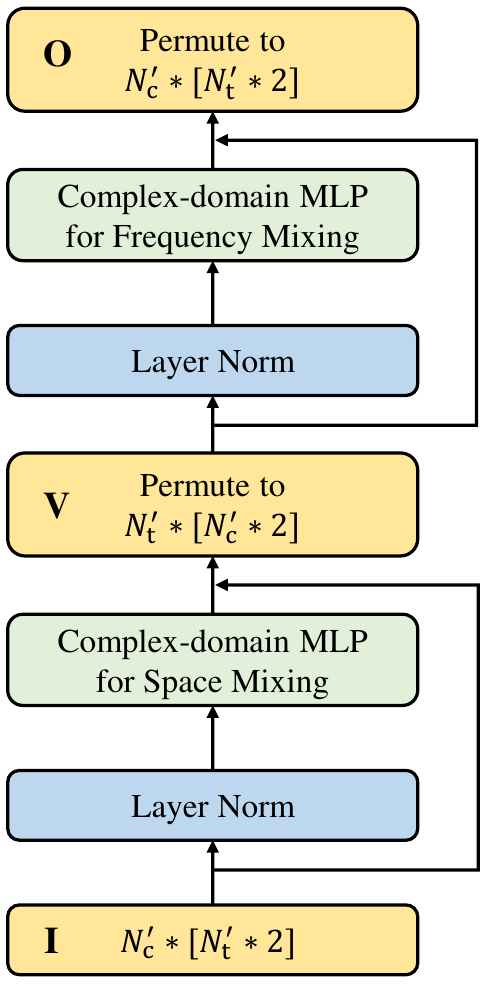}
    \caption*{\footnotesize(c) Interleaved learning module for channel representation}
  \end{minipage}
  \begin{minipage}[t]{0.022\textwidth}
    ~
  \end{minipage}
  \begin{minipage}[t]{0.21\textwidth}
    \centering
    \includegraphics[width=\textwidth]{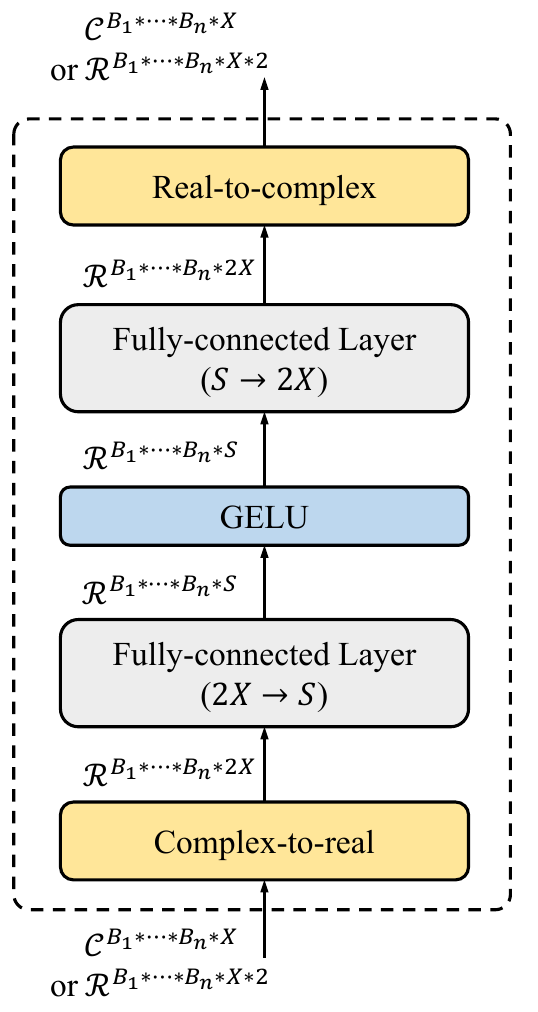}
    \caption*{\footnotesize(d) Complex-domain MLP for Space and Frequency Mixing}
  \end{minipage}
  
  \begin{minipage}[t]{0.0\textwidth}
    ~
  \end{minipage}
  \vspace{0.1em}
  \caption{The graphic illustration of the proposed CMixer model.}
  \vspace{-1em}
  \label{cmixer_fig}
\end{figure*}

\subsection{Interlaced Space and Frequency Characteristics}\label{section3.1}
This subsection presents that there is an unique interlaced space and frequency characteristic in MIMO-OFDM channel. Specifically, according to the channel model in Section \ref{section2.1}, we can define the following function,
\begin{align}
    &{q_{\bf{H}}}(\vec d ,\Delta f) = \nonumber \\ 
    &\sum\limits_{p = 1}^P {{\alpha _p}{e^{ - j2\pi {f_0}{\tau _p}}}{e^{ - j\left( {2\pi {f_0}\vec d  \cdot \vec p /c + 2\pi \Delta f{\tau _p} + 2\pi \Delta f\vec d  \cdot \vec p /c} \right)}}}  ,
\end{align}
where the parameters of ${q_{\bf{H}}}(\cdot)$, $\alpha_p, f_0, \tau_p, {\vec p}$, are the shared large-scale features among all sub-channels. Therefore, ${q_{\bf{H}}}(\cdot)$ is independent of the specific antennas and subcarriers. Then, leveraging ${q_{\bf{H}}}(\cdot)$, the CSI of the $m$-th antenna and $n$-th subcarrier can be written as ${q_{\bf{H}}}({{ \vec d_{m - 1}}} ,\Delta {f_n})$, then the CSI matrix in Eq. \eqref{eq_channel_matrix} can be rewritten as,
{\footnotesize
\begin{align} 
    &{\bf{H}} = \nonumber \\
    &\left[ {\begin{array}{*{20}{c}}
{{q_{\bf{H}}}(\vec 0 ,\Delta {f_1})}&{{q_{\bf{H}}}(\vec 0 ,\Delta {f_2})}& \cdots &{{q_{\bf{H}}}(\vec 0 ,\Delta {f_{{N_{\rm{c}}}}})}\\
{{q_{\bf{H}}}({{ \vec d_1}} ,\Delta {f_1})}&{{q_{\bf{H}}}({{ \vec d_1}} ,\Delta {f_2})}& \cdots &{{q_{\bf{H}}}({{ \vec d_1}} ,\Delta {f_{{N_{\rm{c}}}}})}\\
 \vdots & \vdots & \ddots & \vdots \\
{{q_{\bf{H}}}({{ \vec d_{{N_{\rm{t}}} - 1}}} ,\Delta {f_1})}&{{q_{\bf{H}}}({{ \vec d_{{N_{\rm{t}}} - 1}}} ,\Delta {f_2})}& \cdots &{{q_{\bf{H}}}({{ \vec d_{{N_{\rm{t}}} - 1}}} ,\Delta {f_{{N_{\rm{c}}}}})}
\end{array}} \right].
\label{channel_matrix_interleave}
\end{align}}

For any two columns of ${\bf{H}}\left[ {:,m} \right],{\bf{H}}\left[ {:,n} \right]$ $(\forall m,n \in \left\{ {0, \ldots ,{N_{\rm{c}}-1}} \right\})$, all antennas' CSI of different subcarriers, although ${\bf{H}}\left[ {:,m} \right],{\bf{H}}\left[ {:,n} \right]$ have different frequency shifts $\Delta {f_m}, \Delta {f_n}$, they still share the same space features $\left[ {\vec 0 ,{{ \vec d_1}} , \ldots ,{{ \vec d_{{N_{\rm{t}}} - 1}}} } \right]$ and the same large-scale features ${q_{\bf{H}}}(\cdot,\cdot)$. Similarly, for any two rows of ${\bf{H}}\left[ {m,:} \right],{\bf{H}}\left[ {n,:} \right](\forall m,n \in \left\{ {0, \ldots ,{N_{\rm{t}}-1}} \right\})$, all subcarriers' CSI of different antennas, although there exists different space shift ${{ \vec d_m}},{{ \vec d_n}}$, they still share the same frequency features $[\Delta {f_1},\Delta {f_2}, \ldots ,\Delta {f_{{N_{\rm{c}}}}}]$ and the same large-scale features ${q_{\bf{H}}}(\cdot)$. To summarize, based on the shared large-scale channel features ${q_{\bf{H}}}(\cdot)$, shared space features $\left[ {\vec 0, {{ \vec d_1}}, \ldots, {{ \vec d_{{N_{\rm{t}}} - 1}}} } \right]$ and shared frequency features $[\Delta {f_1},\Delta {f_2}, \ldots ,\Delta {f_{{N_{\rm{c}}}}}]$ interlaced couple, form the internal correlation of the whole CSI matrix.

\subsection{Interleaved Learning Module for Channel Representation}
According to the revealed property, we propose an interleaved space and frequency learning approach for channel representation, as shown in Fig. \ref{cmixer_fig}(a). This approach first learns the space dimension, and this learning process is shared among all subcarriers, named as space mixing. This sharing of space learning process utilizes the sharing of space characteristics among all subcarriers. Then, the proposed approach learns the frequency dimension, and this learning process is also shared among all antennas, called frequency mixing. Similarly, this sharing of frequency learning process also utilizes the sharing of frequency characteristics among all antennas. The cascading of space mixing and frequency mixing interlaces the learned space and frequency characteristics, highly matching the revealed channel characteristics. In the neural network implement, similar to \cite{mlp_mixer}, we also introduce the residual connection and the layer normalization to keep the effectiveness and stability of training. Fig. \ref{cmixer_fig}(c) presents the specific learning module design, and the calculation process is as follows,
\begin{subequations}
\begin{align}
    {\bf{V}}[i,:,:] &= {\bf{I}}[i,:,:] + {\rm{SM}}({\rm{LN}}({\bf{I}}[i,:,:])), \nonumber \\
    &~~~~~~~~~~~\forall i \in \left[ {0,1, \ldots ,N_{\rm{c}}^{\rm{'}} - 1} \right], \\
    {\bf{O}}[:,j,:] &= {\bf{V}}[:,j,:] + {\rm{FM}}({\rm{LN}}({\bf{V}}[:,j,:])), \nonumber \\
    &~~~~~~~~~~~\forall j \in \left[ {0,1, \ldots ,N_{\rm{t}}^{\rm{'}} - 1} \right],
\end{align}
\label{interleaved_equation}
\end{subequations}
where $\bf{I}$ is the input, $\bf{O}$ is the output, $\bf{V}$ is the intermediate variable matrix, $\bf{I}, \bf{O}, \bf{V}$ are all of $N_{\rm{c}}^{'}{\rm{ \times }}{N_{\rm{t}}^{'}}{\rm{ \times 2}}$ size, and $\rm{SM}$, $\rm{FM}$, and $\rm{LN}$ are the abbreviation of space mixing, frequency mixing, and layer normalization, respectively. The computation in Eq. \eqref{interleaved_equation} completes a nonlinear representation of the input $\bf{I}$ through interleaved learning and yields an output $\bf{O}$ of the same size as $\bf{I}$.

Considering that the channel characteristics are embodied in complex-valued data while the current mainstream DL platforms are based on real-valued operations, we design a complex-domain MLP (CMLP) module to specifically realize space mixing and frequency mixing. Fig. \ref{cmixer_fig}(d) shows the flowchart of CMLP, and the simplified calculation process can be written as follows,
\begin{equation}{\mathcal{C}^{X}/\mathcal{R}^{X*2}}\xrightarrow{{{\text{reshape}}}}{\mathcal{R}^{2X}}\xrightarrow{{{\text{MLP}}}}{\mathcal{R}^{2X}}\xrightarrow{{{\text{reshape}}}}{\mathcal{R}^{X*2}}/\mathcal{C}^{X},
\end{equation}
where $\mathcal{C}$ represents complex-valued data 
and $\mathcal{R}$ represents real-valued data, $X$ refers to $N_{\rm{t}}^{'}$ or $N_{\rm{c}}^{'}$. $S$ in Fig. \ref{cmixer_fig}(d) is constant multiples of $X$. This method places the real and imaginary parts in the same neural layer for learning, which ensures the sufficient interaction of real and imaginary information better than treating the real and imaginary parts as separate channels. Moreover, this method recovers the complex dimension at the end to ensure that the next mixing can still be performed on the complex domain. Thus, the learning of the channel data can always remain in the complex domain, even if stacking multiple mixing layers. In summary, this subsection establishes an interleaved learning module, a CMixer layer, consisting of a cascade of a space-learning CMLP and a frequency-learning CMLP. It is suitable for channel representation learning since it is designed according to the channel's physical properties.

\subsection{CMixer Model for Channel Mapping}
By stacking the above representation modules to enhance the nonlinearity and adding the necessary dimension mapping modules,  the CMixer model  is built  for the channel mapping task. The network structure is shown in Fig.~\ref{cmixer_fig}(b), and the calculation process is as follows. 
First, inputting the $N_{\rm{c}}^0{\rm{ \times }}N_{\rm{t}}^0{\rm{ \times 2}}$-size known channel to a $2N_{\rm{t}}^0 \to 2N_{\rm{t}}^{'}$ MLP layer and a $2N_{\rm{c}}^0 \to 2N_{\rm{c}}^{'}$ MLP layer to output a $N_{\rm{c}}^{'}{\rm{ \times }}N_{\rm{t}}^{'}{\rm{ \times 2}}$-size variable. This part yields $\mathcal{O}(N_{\rm{t}}^0N_{\rm{t}}^{'} + N_{\rm{c}}^0N_{\rm{c}}^{'})$ parameters and $\mathcal{O}(N_{\rm{t}}^0N_{\rm{t}}^{'}N_{\rm{c}}^0 + N_{\rm{c}}^0N_{\rm{c}}^{'}N_{\rm{t}}^{'})$ computation complexity. Then, inputting the $N_{\rm{c}}^{'}{\rm{ \times }}N_{\rm{t}}^{'}{\rm{ \times 2}}$-size variable to stacked $K$ CMixer layers and the output variable is still of $N_{\rm{c}}^{'}{\rm{ \times }}N_{\rm{t}}^{'}{\rm{ \times 2}}$ size. This part yields $\mathcal{O}(KN_{\rm{t}}^{'}{S_{\rm{t}}} + KN_{\rm{c}}^{'}{S_{\rm{c}}})$ parameters and $\mathcal{O}(KN_{\rm{t}}^{'}{S_{\rm{t}}}N_{\rm{c}}^{'} + KN_{\rm{c}}^{'}{S_{\rm{c}}}N_{\rm{t}}^{'})$ computation complexity.  Finally, using a $2N_{\rm{t}}^{'}\to 2{N_{\rm{t}}}$ MLP layer and a $2N_{\rm{c}}^{'}\to 2{N_{\rm{c}}}$ MLP layer to mapping the previous output to ${N_{\rm{c}}}{\rm{ \times }}{N_{\rm{t}}}{\rm{ \times 2}}$-size variable, which is the whole MIMO-OFDM CSI. This part yields $\mathcal{O}(N_{\rm{t}}^{'}{N_{\rm{t}}} + N_{\rm{c}}^{'}{N_{\rm{c}}})$ parameters and $\mathcal{O}(N_{\rm{t}}^{'}{N_{\rm{t}}}N_{\rm{c}}^{'} + N_{\rm{c}}^{'}{N_{\rm{c}}}{N_{\rm{t}}})$ computation complexity. 
Since $N_{\rm{t}}^0 < {N_{\rm{t}}},N_{\rm{c}}^0 < {N_{\rm{c}}}$, $N_{\rm{t}}^{'}$ and ${S_{\rm{t}}}$ are constant multiples of $N_{\rm{t}}$, $N_{\rm{c}}^{'}$ and ${S_{\rm{c}}}$ are constant multiples of $N_{\rm{c}}$, the total parameter scale of this model is $\mathcal{O}\left( {KN_{\rm{t}}^2 + KN_{\rm{c}}^2} \right)$ and the total computation complexity is $\mathcal{O}\left( {KN_{\rm{t}}^2{N_{\rm{c}}} + KN_{\rm{c}}^2{N_{\rm{t}}}}\right)$.
To train the model, mean square error (MSE) is used as the loss function, which can be written as
\begin{equation}
   \mathrm{Loss}(\Theta ) = \frac{1}{N}\sum\limits_{num = 1}^N \left\| {{{{\mathbf{H}}}_{num}} - {{\widehat {\mathbf{H}}}_{num}}} \right\|_2^2,
\label{MSE}
\end{equation}
where $\Theta$ is the parameter set of the proposed model, $N$ is the number of channel samples in the training set, and $\Vert \cdot \Vert_2$ is the Euclidean norm. In addition, $\Theta$ can be updated through the existing gradient descent-based optimizers, such as the adaptive momentum estimation (Adam) optimizer.

\section{Simulation Experiments}\label{section4}
In this section, we evaluate the performance and properties of the proposed CMixer through simulation experiments. 

\subsection{Experiment Settings}\label{section4.1}
In this work, we use the open-source DeepMIMO dataset of `O1' scenario at 3.5GHz \cite{deepmimo} for the simulation experiment. The bandwidth is set as 40MHz and divided to 32 subcarriers, the BS is equipped with a uniform linear array (ULA) consisting of 32 antennas. The user area is chose as R701-R1400. 50000 channel samples are randomly sampled from the dataset and divided into the training set and testing set with a ratio of 4:1. 
In addition, for intuitive performance evaluation, we use two existing typical DL-based channel mapping schemes as benchmarks. One is the MLP method \cite{channel_mapping}, using a pure MLP network to learn the mapping function. The other is the Res\_CNN method \cite{res_cnn_es}, using a CNN with residual connection as the network structure.  
Also, the model and training settings are summarized in Table \ref{table_training}.
\begin{table}\footnotesize
	\caption{Parameter settings of CMixer Model}
	\vspace{-0.8em}
	\begin{center}
		\begin{tabular}{ p{2.8cm}   p{4.8cm}}
			\toprule
			\textbf{Parameters} & \textbf{Value} \\
			\toprule
			Batch size & 250\\
			Learning rate &  $1\times10^{-3}$ (multiply by 0.2 every 500 epochs after the 500th epoch)\\
                Model setting & $K=5,  N_{\rm{c}}=32, N_{\rm{t}}=32, N_{\rm{c}}^{'}=32, N_{\rm{t}}^{'}=32, S_{\rm{c}}=128, S_{\rm{t}}=128$\\
                Subset $\Omega {\rm{ = A}} \otimes {\rm{B}}$ & ${\rm{A}}=\{ {0,step_{\rm{t}}, \ldots ,(N_{\rm{t}}^0 - 1) \times step_{\rm{t}}} \}$, ${\rm{B}}=\{ {0,step_{\rm{c}}, \ldots ,(N_{\rm{c}}^0 - 1) \times step_{\rm{c}}} \}$, where $step_{\rm{t}} = \left\lfloor {{N_{\rm{t}}}/N_{\rm{t}}^0} \right\rfloor , step_{\rm{c}} =\left\lfloor {{N_{\rm{c}}}/N_{\rm{c}}^0} \right\rfloor$\\
                Number of model parameters  &  0.176 million, the case of ($N_{\rm{c}}^0 =5, N_{\rm{t}}^0=5$) \\
                Floating-point operations (FLOPs) & 5.61 million, the case of ($N_{\rm{c}}^0 =5, N_{\rm{t}}^0=5$)  \\
			Training epochs & 2000 \\			
			Optimizer & Adam \\
			\toprule
		\end{tabular}		
	\end{center}
        \vspace{-1.8em}
	\label{table_training}
\end{table}
Moreover, we use normalized MSE (NMSE) and cosine correlation $\rho$ \cite{csinet} as the performance indices, which are defined as follows:
\begin{equation}
    \mathrm{NMSE} = \mathbb{E}\left\{ {\frac{{\left\| {{\mathbf{H}} - \widehat {\mathbf{H}}} \right\|_2^2}}{{\left\| {\mathbf{H}} \right\|_2^2}}} \right\},
\label{NMSE}
\end{equation}
and
\begin{equation}
    \rho {\rm{ = }}{\mathop{\mathbb{E} }\nolimits} \left\{ {\frac{1}{{{N_c}}}\sum\limits_{m = 1}^{{N_c}} {\frac{{|{{\widehat {\bf{h}}}_m}^H{{\bf{h}}_m}|}}{{||{{\widehat {\bf{h}}}_m}|{|_2}||{{\bf{h}}_m}|{|_2}}}} } \right\},
\label{rou}
\end{equation}
where the ${\bf{h}_m}/{\widehat {\bf{h}}}_m$ is the CSI of $m$-th subcarrier, i.e. the $m$-th columon of the CSI matrix $\mathbf{H}/\widehat {\mathbf{H}}$.

\subsection{Performance Evaluation} \label{section3.2}
We first evaluate the mapping performance of the proposed CMixer and use various mapping settings to ensure the universality of evaluation results. The known channel size, the $\Omega$ in Eq. \eqref{eq_mapping}, is set as 4×4, 5×5, 6×6 and 7×7 (antennas × subcarriers). Table \ref{table_performance} shows the mapping performance of proposed scheme and benchmarks. Compared with benchmarks, the proposed CMixer provides 4.6\textasciitilde10dB gains in NMSE and 5.7\textasciitilde9.6 percentage points gain in $\rho$ under same known channel size, or reduces required known channel size by up to $67.3\%(1-16/49)$ under same and even better accuracy.

\begin{table}[htbp]\centering\footnotesize
\renewcommand{\arraystretch}{1.2}
\caption{NMSE and $\rho$ under various mapping settings.}
\begin{tabular}{>{\centering\arraybackslash}p{0.95cm}|>{\centering\arraybackslash}p{1.35cm}|cccc}
\hline
\multirow{2}{*}{\textbf{Indexes}} & \multirow{2}{*}{\textbf{Schemes}} & \multicolumn{4}{c}{\textbf{\makecell{Known channel size\\(antennas × subcarriers)}}}           
\\ \cline{3-6} 
 &          & \multicolumn{1}{c|}{4 × 4} & \multicolumn{1}{c|}{5 × 5} & \multicolumn{1}{c|}{6 × 6} & 7 × 7 \\ \hline
\multirow{3}{*}{\makecell{NMSE\\(dB)}}    & CMixer               & \multicolumn{1}{c|}{\bf{-10.49}} & \multicolumn{1}{c|}{\bf{-14.11}} & \multicolumn{1}{c|}{\bf{-15.38}} & \bf{-19.24} \\ \cline{2-6} 
 & MLP      & \multicolumn{1}{c|}{-5.85}    & \multicolumn{1}{c|}{-6.88}    & \multicolumn{1}{c|}{-7.53}    &   -9.22  \\ \cline{2-6} 
 & Res\_CNN & \multicolumn{1}{c|}{-3.37}    & \multicolumn{1}{c|}{-4.49}    & \multicolumn{1}{c|}{-7.23}    & -8.94   \\ \hline
\multirow{3}{*}{$\rho$}       & CMixer               & \multicolumn{1}{c|}{\bf{0.9555}} & \multicolumn{1}{c|}{\bf{0.9807}} & \multicolumn{1}{c|}{\bf{0.9856}} & \multicolumn{1}{c}{\bf{0.9940}} \\ \cline{2-6} 
 & MLP      & \multicolumn{1}{c|}{0.8589}    & \multicolumn{1}{c|}{0.8903}    & \multicolumn{1}{c|}{0.9071}    &   0.9367  \\ \cline{2-6} 
 & Res\_CNN & \multicolumn{1}{c|}{0.7394}    & \multicolumn{1}{c|}{0.8079}    & \multicolumn{1}{c|}{0.9017}    &   0.9353  \\ \hline
\end{tabular}
\label{table_performance}
\end{table}

\begin{figure}[htbp]
\centering
  \includegraphics[width=0.48\textwidth]{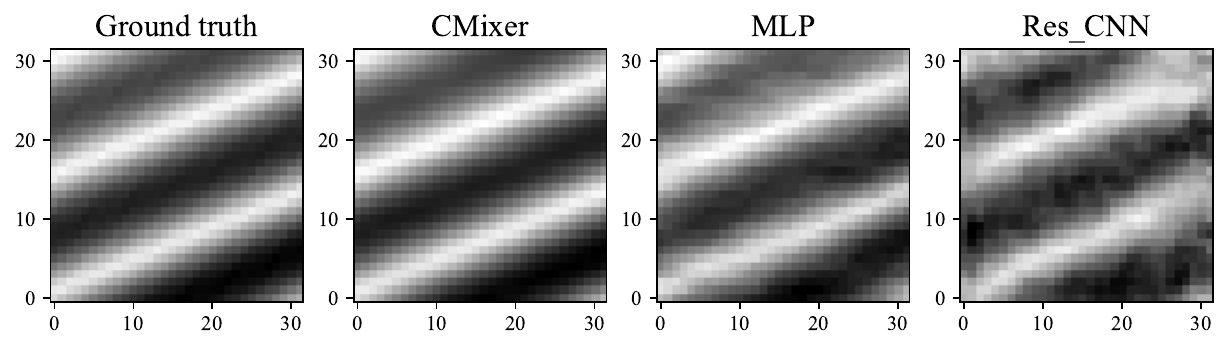}
      \caption{Grayscale visualization of the original CSI and represented CSI.}
     \vspace{-0.8em}
  \label{visual_fig}
  \vspace{-0.5em}
\end{figure}

Further, Fig. \ref{visual_fig} shows the mapping results of the different schemes on a random channel sample using grayscale visualization (Known channel size is 5 antennas × 5 subcarriers). It can be intuitively seen that the mapping channel based on CMixer is closer to the true channel than based on benchmarks. In summary, the above experimental results reflect the effectiveness and superiority of the proposed CMixer on the channel mapping task.

\subsection{Ablation Studies}

Complex-domain computation and interleaved learning are the two typical features of the proposed CMixer model. The performance evaluation in Section \ref{section3.2} illustrates the superiority of the scheme as a whole, and this subsection evaluates the value of these sub-modules through ablation studies. These ablation experiments are all conducted under the known channel of 5 antennas × 5 subcarriers.

We first evaluate the necessity of using CMLPs. We replace the proposed CMLPs with vanilla MLPs. The specific ablation operation is changing the $2X \to S \to 2X$ calculation (CMLP) to the $2\left( {X \to S \to X} \right)$ calculation (MLP with real and imag parallelism), remaining the same computation complexity, $4XS$, to ensure fairness. Table \ref{table_ablation1} shows the results of this ablation. Not using the CMLPs in either space mixing or frequency mixing leads to performance degradation. This phenomenon demonstrates the advantages of CMLPs in the representation learning of complex-valued channel data.

\vspace{-0.3em}
\begin{table}[htbp]\footnotesize
\caption{{NMSE (dB) of using/not using CMLPs.}}
\vspace{-0.6em}
\renewcommand{\arraystretch}{0.75}
\begin{center}
    \begin{tabular}{c|c|c}
\hline
\textbf{\diagbox{\makecell{Frequency\\Mixing}}{\makecell{Space\\Mixing}}}   & { Using MLP} & { Using CMLP} \\ \hline
\multirow{2}{*}{ Using MLP} &     \multirow{2}{*}{ -7.13}  &     \multirow{2}{*}{ -11.21} \\ & & \\ \hline
\multirow{2}{*}{ Using CMLP}  &    \multirow{2}{*}{ -12.38}    &     \multirow{2}{*}{{  \bf{-14.11}}} \\ & & \\ \hline
\end{tabular}
\end{center}
\label{table_ablation1}
\vspace{-0.8em}
\end{table}

Then, we also evaluate what leads to the effectiveness of the proposed interleaved learning and whether it exactly stems from the unique interlaced space and frequency characteristics in the channel. 
Specifically, we shuffle the element order in the channel matrix in a way that remains/loses interlaced space and frequency characteristics and use the CMixer to learn the new matrix as the representation output. The first shuffle mode is named as `interlaced shuffle' and written as,
\begin{equation}{
    {{\bf{H}}_{{\rm{ interlaced}}}} = {{\bf{P}}^{{N_{\rm{t}}} \times {N_{\rm{t}}}}}  {{\bf{H}}^{{N_{\rm{t}}} \times {N_{\rm{c}}}}}  {{\bf{P}}^{{N_{\rm{c}}} \times {N_{\rm{c}}}}},}
\end{equation}
where each $\bf{P}$ represents a permutation matrix. In this way, the new channel matrix ${{\bf{H}}_{{\rm {interlaced}}}}$ has different space and frequency characteristics from the original channel ${\bf{H}}$, but its space and frequency characteristics are still interlaced coupled since its each row or column still respectively corresponds to certain antenna or subcarrier. The other shuffle mode is named as `non interlaced shuffle' and written as,
\begin{equation}{
    {{\bf{H}}_{{\rm{non}} - {\rm{interlaced}}}} = {\rm{vec}}^{ - 1}\left( {{{\bf{P}}^{{N_{\rm{t}}}{N_{\rm{c}}} \times {N_{\rm{t}}}{N_{\rm{c}}}}} {\rm{vec}}\left( {{{\bf{H}}^{{N_{\rm{t}}} \times {N_{\rm{c}}}}}} \right)} \right),}
\end{equation}
where ${\rm{vec}}(\cdot)$ denotes the transposition of a matrix into a vector and ${\rm{vec}}^{ - 1}(\cdot)$ is the reverse process of ${\rm{vec}}(\cdot)$. In this shuffle mode, the row or column of ${{\bf{H}}_{{\rm{non}} - {\rm{interlaced}}}}$ does not necessarily correspond to certain antenna or subcarrier. Thus, the space and frequency characteristics are no longer coupled in an interlaced manner. Note that the above two kinds of new channel matrix ${{\bf{H}}_{{\rm{non}} - {\rm{interlaced}}}}, {{\bf{H}}_{{\rm {interlaced}}}}$ and the original channel matrix ${\bf{H}}$ are identical in element content and only differ on the data structure. We use various random permutation matrices for experiments, and the statistical results are shown in Table \ref{table_ablation2}. 
The proposed CMixer can effectively represent ${{\bf{H}}_{{\rm{ interlaced}}}}$ with interlaced space and frequency characteristics while is severely degraded in representing ${{\bf{H}}_{{\rm{non}} - {\rm{interlaced}}}}$ without the interlaced characteristics. This phenomenon illustrates that the excellent performance of the proposed interleaved space and frequency learning indeed draws on the physical properties rather than relying exclusively on data fitting. The CMixer model's performance gains come from the design inspired by channel characteristics.

\begin{table}[htbp]\footnotesize
\vspace{-0.3em}
\renewcommand{\arraystretch}{1.2}
	\caption{\centering NMSE (dB) of using CMixer to map channel w/wo interlaced space and frequency characteristics.}
	\vspace{-0.8em}
	\begin{center}
		\begin{tabular}{ p{3cm}   p{3.5cm}}
                \hline
			\textbf{Shuffle mode} & \textbf{NMSE (dB)} \\  \hline
                Origin & $-14.11$ \\ 
                Interlaced       &  $-14.25\pm0.20$ \\ 
			Non-Interlaced & $-1.01\pm0.05$ \\ \hline
		\end{tabular}		
	\end{center}
	\vspace{-0.9em}
	\label{table_ablation2}
\end{table}

\section{Conclusion} \label{conclusion}
In this paper, we propose a CMixer model for channel mapping, achieving excellent performance. Modeling analysis and ablation studies state that the complex-domain computation and interleaved learning in this model are suitable for channel representation and, thus, are likely to be applicable in other channel-related tasks as well. We hope that our work can provide inspirations for using DL technologies to solve problems in wireless systems.

\ifCLASSOPTIONcaptionsoff
  \newpage
\fi

\bibliographystyle{IEEEbib}
\bibliography{ref}

\begin{thebibliography}{1}

\bibitem{bigAI6G}
Z.~Chen, Z.~Zhang, and Z.~Yang,
\newblock ``Big {AI} models for {6G} wireless networks: Opportunities, challenges, and research directions,''
\newblock {\em arXiv preprint arXiv:2308.06250}, 2023.

\bibitem{channel_mapping}
M.~Alrabeiah and A.~Alkhateeb,
\newblock ``Deep learning for {TDD} and {FDD} massive {MIMO}: Mapping channels in space and frequency,''
\newblock in {\em 2019 53rd Asilomar Conference on Signals, Systems, and Computers}, 2019, pp. 1465--1470.

\bibitem{channelesnet}
M.~Soltani, V.~Pourahmadi, A.~Mirzaei, et~al.,
\newblock ``Deep learning-based channel estimation,''
\newblock {\em IEEE Communications Letters}, vol. 23, no. 4, pp. 652--655, 2019.

\bibitem{res_cnn_es}
L.~Li, H.~Chen, H.-H. Chang, et~al.,
\newblock ``Deep residual learning meets {OFDM} channel estimation,''
\newblock {\em IEEE Wireless Communications Letters}, vol. 9, no. 5, pp. 615--618, 2020.

\bibitem{2Dseq2seq}
Z.~Chen, Z.~Zhang, Z.~Xiao, et~al.,
\newblock ``Viewing channel as sequence rather than image: A {2-D} {Seq2Seq} approach for efficient {MIMO-OFDM} {CSI} feedback,''
\newblock {\em IEEE Transactions on Wireless Communications}, vol. 22, no. 11, pp. 7393--7407, 2023.

\bibitem{mlp_mixer}
I.O. Tolstikhin, N.~Houlsby, A.~Kolesnikov, et~al.,
\newblock ``{MLP-Mixer}: An all-{MLP} architecture for vision,''
\newblock {\em Advances in Neural Information Processing Systems}, vol. 34, pp. 24261--24272, Dec. 2021.

\bibitem{deepmimo}
A.~Alkhateeb,
\newblock ``{DeepMIMO}: A generic deep learning dataset for millimeter wave and massive {MIMO} applications,''
\newblock {\em arXiv preprint arXiv:1902.06435}, 2019.

\bibitem{csinet}
C.-K. Wen, W.-T. Shih, and S.~Jin,
\newblock ``Deep learning for massive {MIMO} {CSI} feedback,''
\newblock {\em IEEE Wireless Communications Letters}, vol. 7, no. 5, pp. 748--751, Oct. 2018.

\end{thebibliography}
\end{document}